\documentclass[10pt, conference]{ieeeconf}  %

\IEEEoverridecommandlockouts
\usepackage{graphicx}
\usepackage{xcolor}
\usepackage[caption=false,font=footnotesize]{subfig}
\usepackage{url}
\usepackage{cite}

\usepackage{array, amsmath, amssymb, amsfonts, bm}
\usepackage{mathrsfs}
\usepackage{multirow, booktabs}
\usepackage{algorithm, algpseudocode}
\usepackage{xargs}

\usepackage{caption}
\usepackage{comment}
\usepackage{kantlipsum}
\usepackage{todonotes}
\usepackage{microtype}

\newcommand{\argmax}{\operatornamewithlimits{argmax}}

\newcommand{\setR}{\mathbb{R}}
\newcommand{\setRp}{\mathbb{R}_+}
\newcommand{\setC}{\mathbb{C}}
\newcommand{\setSp}{\mathbb{S}_+}

\newcommand{\eye}{{\bf I}}

\newcommand{\T}{\mathsf{T}}

\newcommand{\distnormal}[2]{\mathcal{N}\left({#1}, {#2}\right)}
\newcommand{\distcmpnormal}[2]{\mathcal{N}_{\mathbb{C}}\left({#1}, {#2}\right)}

\newcommand{\E}{\mathbb{E}}

\NewDocumentCommand\newletter{m m o m m}{%
\NewDocumentCommand#1{s t@ o}{%
\IfBooleanTF{##1}{\mathbf{\MakeUppercase{#2}}\IfValueT{#3}{^{#3}}}{%
\IfBooleanTF{##2}{\mathbf{#2}\IfValueT{#3}{^{#3}}_{\IfValueTF{##3}{##3}{#5}}}{%
{#2}\IfValueT{#3}{^{#3}}_{\IfValueTF{##3}{##3}{#4}}%
}}}}

\NewDocumentCommand\newletterbm{m m o m m}{%
\NewDocumentCommand#1{s t@ o}{%
\IfBooleanTF{##1}{\bm{\MakeUppercase{#2}}\IfValueT{#3}{^{#3}}}{%
\IfBooleanTF{##2}{\bm{#2}\IfValueT{#3}{^{#3}}_{\IfValueTF{##3}{##3}{#5}}}{%
{#2}\IfValueT{#3}{^{#3}}_{\IfValueTF{##3}{##3}{#4}}%
}}}}

\newletter{\y}{y}{ft}{f}
\newletter{\x}{x}{}{}

\newletter{\s}{s}{k,ft}{k}
\newletter{\e}{n}{ft}{f}

\newletter{\w}{w}{k,lf}{k,l}
\newletter{\h}{h}{k,lt}{k,l}

\newletter{\we}{w}[\mathrm{noi}]{k,lf}{k,l}
\newletter{\he}{h}[\mathrm{noi}]{k,lt}{k,l}

\newletterbm{\feat}{\mu}{k}{}
\newletter{\ests}{\hat{s}}{k,ft}{k}

\title{\LARGE \bf
Infrastructure-less Localization from Indoor Environmental Sounds \\ Based on Spectral Decomposition and Spatial Likelihood Model
}

\author{Satoki Ogiso$^{1}$, Yoshiaki Bando$^{1}$, Takeshi Kurata$^{1}$ and Takashi Okuma$^{1}$%
\thanks{$^{1}$National Institute of Advanced Industrial Science and Technology (AIST), Japan
        {\tt\small \{s.ogiso, y.bando, t.kurata, takashi-okuma\}@aist.go.jp}}%
}

\usepackage{eso-pic,xcolor}
\AddToShipoutPictureBG*{%
  \put(\LenToUnit{.1\paperwidth},\LenToUnit{3em})%
    {\parbox[3pt]{49.5em}{%
      \footnotesize\copyright 2023 IEEE.  Personal use of this material is permitted.  Permission from IEEE must be obtained for all other uses, in any current or future media, including reprinting/republishing this material for advertising or promotional purposes, creating new collective works, for resale or redistribution to servers or lists, or reuse of any copyrighted component of this work in other works.
      }%
    }%
}

\begin{document}
\bstctlcite{IEEE:BSTcontrol}

\maketitle
\thispagestyle{empty}
\pagestyle{empty}

\begin{abstract}
Human and/or asset tracking using an attached sensor units helps understand their activities.
Most common indoor localization methods for human tracking technologies require expensive infrastructures, deployment and maintenance.
To overcome this problem, environmental sounds have been used for infrastructure-free localization.
While they achieve room-level classification, they suffer from two problems: low signal-to-noise-ratio (SNR) condition and non-uniqueness of sound over the coverage area.
A microphone localization method was proposed using supervised spectral decomposition and spatial likelihood to solve these problems.
The proposed method was evaluated with actual recordings in an experimental room with a size of 12$\times$30~m.
The results showed that the proposed method with supervised NMF was robust under low-SNR condition compared to a simple feature (mel frequency cepstrum coefficient: MFCC).
Additionally, the proposed method could be easily integrated with prior distribution, which is available from other Bayesian localizations.
The proposed method can be used to evaluate the spatial likelihood from environmental sounds.
\end{abstract}

\section{Introduction}

Human and/or asset localization with attached sensor units is essential to understanding their activities and enhancing workspace productivity.
Since indoor localization systems are installed at the expense of the beneficiaries, a number of localization methods are available according to requirements.
A common method for measuring human behavior is to install devices that emit signals at known locations.
This method achieves up to centimeter-grade accuracy, but the installation and maintenance of these devices are costly.
Thus, the importance of the infrastructure-less localization system lies in its simple deployment.

Infrastructure-less localization has been mostly studied using existing radio waves or electromagnetic fields in the environment.
Camera-based localization achieves high accuracy while consuming energy; it is sometimes prohibited in facilities for security reasons.
A major approach is to use the signal strength of existing Wi-Fi base stations.
This technology requires many base stations, so it is limited to large facilities with Wi-Fi, such as a train station.
Using magnetic anomalies, precise localization within a few meters of area is possible. 
Similar anomalies may occur in every few meters, because the location-specific cue is simply a three-dimensional vector of magnetometer.
Dead reckoning with IMU is completely independent of environment, whereas the absolute location needs to be provided with other methods.

This study focuses on infrastructure-less localization based on location-specific environmental sounds.
Environmental sounds exist in most of the facilities and contain more information than magnetic fields.
Some methods record sounds in advance to train classifiers for room or area classifiers.
However, use of these sounds has two challenges: low-SNR conditions and non-uniqueness over a coverage area.
Low-SNR signals are not similar to training samples, so classifiers fail.
In addition, environmental sounds may not be unique at several locations, such as fans of the same type in each corridor.
To solve this problem, environmental sounds are only used for coarse localization as room/area classification.
It would be useful for infrastructure-free indoor localization if position with coordinate from environmental sound at low SNR could be estimated.

\begin{figure}[t] %
    \centering
    \vspace{2mm}
    \subfloat[Na\"ive direct regression-based localization]{\includegraphics{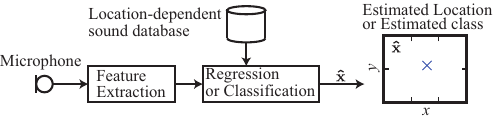}} \\[3mm]
    \subfloat[Proposed spatial likelihood-based localization]{\includegraphics{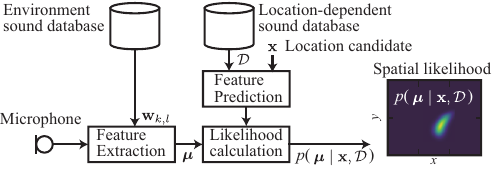}}
    \caption{Comparison between na\"ive direct regression-based localization and our spatial likelihood-based localization.}
    \label{fig:bg}
    \vspace{-4mm}
\end{figure}

\begin{figure*}[!t]
    \centering
    \vspace{1mm}
    \includegraphics[width=\hsize]{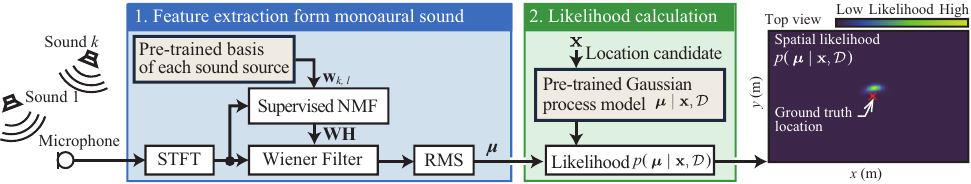}
    \caption{Architecture of the proposed localization method.}
    \label{fig:details}
    \vspace{-4mm}
\end{figure*}

In this study, a sound-based infrastructure-less localization method is proposed that can deal with low-SNR conditions and ill-conditioned measurements that cannot determine a unique location.
The comparison between conventional and proposed methods is shown in Fig.~\ref{fig:bg}.
This study addresses the issues listed above. 
The proposed method first extracts source root-mean-square (RMS) values from an observed mixture signal using supervised non-negative matrix factorization (NMF)~\cite{fevotte2008nonnegative} and a Wiener filter.
The supervised NMF and Wiener filter reduce out-of-domain noise, by decomposing the sound into components which corresponds to the premeasured environmental sounds.
Then spatial likelihood is calculated by comparing them with predicted distribution from a Gaussian process (GP) regression model.
GP provides a sound model in the environment as probabilistic distribution by combining with the supervised NMF.
The proposed method enables maximum a posteriori (MAP) estimation with any Bayes-based indoor localization.

\section{Related Work}
\subsection{Infrastructure-assisted methods}
The majority of the indoor localization methods require special infrastructure in the environment.
The most common method uses the received signal strength indicator (RSSI) of Bluetooth low energy (BLE) or radio-frequency identifier (RFID)~\cite{bahlRADARInbuildingRFbased2000}, which achieves 2--3~m accuracy~\cite{murataSmartphonebasedIndoorLocalization2018,uranoEndtoEndBLEIndoor2021}.
The major drawback of these methods is that they require numerous BLE beacons to be deployed.
For example, a case study in a five-story building deployed 218 beacons~\cite{murataSmartphonebasedLocalizationBlind2019}.
Other infrastructure-based methods such as ultra-wideband (UWB) or ultrasound~\cite{urenaAcousticLocalPositioning2018,ogisoRobustAcousticLocalization2021a} require wiring between base stations to achieve accurate localization. 
Installing and removing of these beacons or base stations are costly.
Moreover, maintaining the health of these beacons, such as battery life or fault, is exhausting work.

\subsection{Infrastructure-less methods}
To address this challenge, infrastructure-less indoor localization methods have been proposed. 
These methods use existing environmental cues or self-contained sensors.
A camera~\cite{campos2021orb} or light detection and ranging (LiDAR) sensor~\cite{labbeRTABMapOpensourceLidar2019} achieves the most accurate localization.
However, these sensors have a high power requirement, and are sometimes not allowed for security reasons.
WiFi-based methods~\cite{bahlRADARInbuildingRFbased2000} use Wi-Fi base stations that are already installed as beacons.
However, several Wi-Fi stations must be simultaneously detected, which is not possible in Wi-Fi-restricted locations such as factories.
Another useful environmental cue is the local magnetic anomalies caused by the ferromagnetic materials in a building~\cite{xieMaLocPracticalMagnetic2014}.
These magnetic anomalies are unique within a few meters, but not throughout the entire building. 
This is due to the fact that a magnetic anomaly is merely a three-dimensional vector that can be almost identical in every few meters.
Another example is IMU-based indoor localization, which uses the estimated gravity vector, linear velocities, and angular velocities~\cite{kourogiPersonalPositioningBased2003}.
They use specific acceleration pattern~\cite{kourogiPersonalPositioningBased2003} or zero-velocity data~\cite{foxlinPedestrianTrackingShoemounted2005}.
This method achieves high short-term accuracy, but absolute localization is required for long term use.
Since these methods have different advantages and disadvantages, the Bayesian integration of these sensors has also been actively researched~\cite{yangSurveyRecentIndoor2021,kimIndoorPositioningSystem2017}.%

\subsection{Environmental sound-based methods}
Environmental sounds are another potential location-specific cue promising for sensor localization.
Environmental sound localization can also be divided into two methods: direction of arrival (DoA)-based methods and fingerprint-based methods.
The DoA-based methods use a microphone array to estimate the DoA of sound sources with sub-meter accuracy~\cite{eversAcousticSLAM2018,rypkemaPassiveInvertedUltraShort2019}.
However, these methods require a microphone array with sufficiently large microphone spacing~\cite{ogisoRobustIndoorLocalization2019}.
The fingerprint-based methods first samples the sounds in the locating area and then use classification or regression to estimate the location of the microphone ~\cite{azizyanSurroundSenseMobilePhone2009,tarziaIndoorLocalizationInfrastructure2011b,aonoInfrasonicSceneFingerprinting2017a}.
The advantages of the sound fingerprint-based methods are similar to those of the infrastructure-less magnetic methods.
In addition, they are more promising than the magnetic methods because the audio spectrum has far more information (usually hundreds of dimensions) than the three-dimensional magnetic information.
However, the fingerprint-based methods have two main problems.
The first is that they ignore individual sound sources and simply use the features of mixed sound samples~\cite{leonardoFrameworkInfrastructureFreeIndoor2018a}.
Since mixed signals vary drastically according to the recording locations, they may be suceptible to the out-of-domain problem.
The second problem is that they only solve classification or regression problems to obtain a single estimation value.
The environmental sounds may not be unique at several locations, such as fans of the same type in each corridor.
The conventional methods cannot afford to represent this multi-modal likelihood as a regression result.
The proposed method estimates spatial likelihood based on environmental sound, and it is compatible with the Bayesian estimations mentioned above.

\section{Proposed Method}

As shown in Fig.~\ref{fig:details}, the proposed method consists of two major steps: NMF-based decomposition step and GP-based likelihood calculation step.
This two-stage framework leads to the high interpretability of the inference and enables us to easily integrate other modalities in a Bayesian manner.

\subsection{Feature extraction with noise-aware supervised NMF}

To extract spectral features from an observed mixture signal, the mixture signal with supervised NMF was decomposed.
Let $\y \in \setC$ be the observed mixture in the time-frequency domain obtained by the short-time Fourier transform (STFT), where $f=1, \ldots, F$ and $t=1, \ldots, T$ are the frequency and time indices, respectively.
A mixture signal $\y$ was assumed to consist of $K$ landmark source signals $\s \in \setC$ and a noise signal $\e \in \setC$ as follows:
\begin{align}
    \y = \sum_{k=1}^K \s + \e \label{eq:mix}
\end{align}
As described later, NMF decomposes the mixture $\y$ into source signals $\s$ whose spectral patterns are known in advance and residual noise (e.g., non-environmental sounds).
The feature vector for localization $\feat@ = [\feat[1], \ldots, \feat[K]] \in \setR^K$ is then calculated as log-RMS values of source estimates $\ests$ as follows:
\begin{align}
    \feat = \frac{1}{2}\log \left( \sum_{f=1}^F \sum_{t=1}^T |\ests|^2 \right). \label{eq:feature}
\end{align}

To extract the source signals from a mixture, we utilize supervised NMF~\cite{bisot2016supervised,fevotte2008nonnegative}.
NMF assumes that each source signal $\s$ follows a local Gaussian model (LGM)~\cite{fevotte2008nonnegative} whose power spectral density is represented by $L_k$ spectral basis (template) vectors $\w@ = [\w[k,1f], \ldots, \w[k,{L_k}f]] \in \setRp^F$ and their temporal activations $\h@ = [\h[k,1t], \ldots, \h[k,{L_k}t]] \in \setRp^T$ as follows:
\begin{align}
    \s &\sim \distcmpnormal{0}{\sum_{l=1}^{L_k} \w \h}, \label{eq:source}
\end{align}
where $\distcmpnormal{\mu}{\sigma^2}$ is a complex Gaussian distribution with a mean parameter $\mu$ and a variance parameter $\sigma^2$.
The basis vectors of each source are obtained in advance from isolated recordings of the landmark sound sources by a maximum likelihood estimation of Eq.\eqref{eq:source}~\cite{fevotte2008nonnegative}.

In the localization phase, an NMF-based LGM on the noise signal $\e$ was assumed with unknown basis vectors $\w@[0,l] = [\w[0,1f], \ldots, \w[0,{L_0}f]] \in \setRp^F$ and their activations  $\h@[0,l] = [\h[0,1t], \ldots, \h[0,{L_0}t]] \in \setRp^T$ as:
\begin{align}
    \e \sim \distcmpnormal{0}{\sum_{l=1}^{L_0} \w[0,lf] \h[0,lt]}. \label{eq:noise}
\end{align}
The unknown parameters $\w@[0,l]$ and $\h@$ ($k=0, \ldots, K$) were estimated to maximize a likelihood function of the mixture signal.
From Eqs.~\eqref{eq:mix}, \eqref{eq:source}, and \eqref{eq:noise}, the likelihood function is derived as follows:
\begin{align}
    \y \sim \distcmpnormal{0}{\sum_{k=0}^K \sum_{l=1}^{L_k} \w \h}.
\end{align}
The estimation of the model parameters is performed with 
multiplication update rules~\cite{fevotte2008nonnegative} as in the original NMF.

Once the model parameters are estimated, the landmark source signals can be estimated by Wiener filtering as follows:
\begin{align}
    \hspace{-3mm}\ests &= \E[\s \mid \y, \w*, \h*] \nonumber \\
    &= \frac{\sum_{l=1}^{L_k} \w \h}{\sum_{k=0}^K \sum_{l=1}^{L_k} \w \h}\y. \label{eq:landmark}
\end{align}

\subsection{Spatial likelihood-based localization with GP regression}
The indoor sound propagation process depends on many factors, such as the direction of sources, reflections, and reverberations.
Thus, sound distribution in the target area for localization was represented by using GP regression~\cite{rasmussen2003gaussian}.
Note that conventional methods use regression (or classification) to estimate the source location directly.
In contrast, the proposed method predicts sound distribution by GP regression and calculates the spatial likelihood.

The GP regression was trained by using mixture recordings $\y*_n$ sampled at locations $\x@[n] \in \setR^2$ ($n = 1, \ldots, N$).
The mixture recordings were initially converted into NMF-based feature values $\feat[n,k]$ for source $k$ at a sampled location $\x@[n]$.
The Gaussian process is then trained to predict the feature values conditioned by the sampled locations $\x* = [\x@[1], \ldots, \x@[N]]$:
\begin{align}
    [\feat[1,k], \ldots, \feat[N,k]]^\T \mid \x* \sim \distnormal{\bm{0}}{\mathbf{K}(\x*) + \sigma^2 \eye}, \label{eq:gp}
\end{align}
where $\sigma^2 \in \setRp$ is a variance hyperparameter, and $\mathbf{K}_k(\x*) \in \setSp^{N\times N}$ is a covariance matrix whose $i,j$-element $\kappa_k(\x@[i],\x@[j])$ is defined by a kernel function.
In this study, we use the scaled radial basis function (RBF) kernel~\cite{rasmussen2003gaussian} for $\kappa_k(\x@[i],\x@[j])$:
\begin{align}
    \kappa_k(\x@[i], \x@[j]) = \theta _k\exp\left(-\frac{\| \x@[i] - \x@[j] \|^2}{2 \gamma_k^2} \right), 
\end{align}
where $\theta_k \in \setRp$ and $\gamma_k \in \setR$ are the kernel parameters.
The training optimizes these parameters to maximize the log-likelihood of Eq.~\eqref{eq:gp} by using gradient descent.

Once the kernel parameters are trained, a likelihood function can be determined for an arbitrary location $\x@$ with an unseen feature vector $\feat@ = [\feat[1], \ldots, \feat[K]]$ by GP regression as follows:
\begin{align}
    \feat@ \mid \x@, \mathcal{D} \sim \prod_{k=1}^K \distnormal{\feat[k] \mid  \hat{m}_{k}(\x@)}{\hat{v}_{k}(\x@)}
\end{align}
where $\mathcal{D} = \{\feat@[1], \x@[1]\ldots, \feat@[N], \x@[N]\}$ are the training data, and $\hat{m}_{k}(\x@) \in \setR$ and $\hat{v}_{k}(\x@) \in \setRp$ are the predictive mean and variance of a feature $\feat$ determined from the pretrained GP model~\cite{rasmussen2003gaussian}.
The microphone location $\hat{\bm{x}}_\mathrm{ML}$ can be estimated by finding the location where the likelihood function for its feature vector is maximized:
\begin{align}
    \hat{\mathbf{x}}_{\rm ML} = \argmax_{\x@} p(\feat@ \mid \x@, \mathcal{D}).
\end{align}
A simple way to find the maximum estimate is to perform a grid search in the likelihood space.
In addition, since our localization is based on a spatial probabilistic (likelihood) model, it is easy to combine with other modalities by introducing a prior distribution.
If a prior distribution $p(\x@)$ is available (e.g., inertial sensor estimate), we can calculate the posterior distribution $p(\x@ \mid \feat@, \mathcal{D}) \propto  p(\feat@ \mid \x@, \mathcal{D})p(\x@)$, and the location can be estimated using maximum-a-posteriori estimation as follows:
\begin{align}
    \hat{\mathbf{x}}_{\rm MAP} = \argmax_{\x@} p(\feat@ \mid \x@, \mathcal{D}) p(\x@). \label{eq:map}
\end{align}

\section{Experimental Evaluation}
\subsection{Experimental setting}
\begin{figure}[!t]
    \centering
    \vspace{1mm}
    \includegraphics{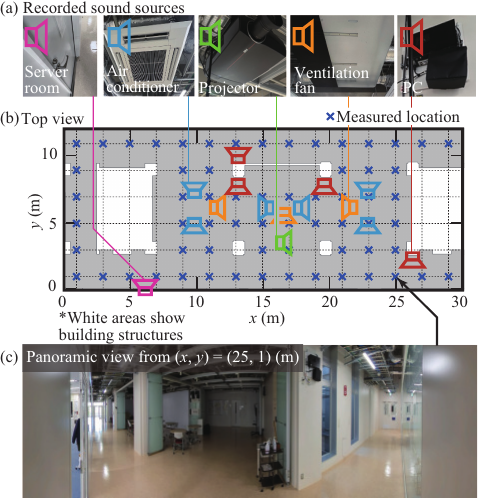}
    \caption{Target area for localization. (a) lists landmark sound sources used in our experiment, (b) shows the top view of the area, and (c) shows a panoramic view of the area.}
    \label{experiment_map}
\end{figure}
The proposed method was evaluated in the indoor environment, as shown in Fig.~\ref{experiment_map}.
The shape of the environment was approximately $(x, y) = (30, 12)$ (m), with pillars and mechanical voids.
This environment has $K=5$ kinds of significant sound sources.
The approximate locations and representative pictures of these sounds are shown in Fig.~\ref{experiment_map}(a).
The sound signals were recorded with a linear PCM recorder (PCM-D100, Sony) and a microphone (M50, EarthWorks) at a sampling frequency of 48 kHz.
To train the GP model and evaluate the proposed method, mixture recordings were sampled on a grid with 2~m spacing and a height of $z=1$~m, as shown in Fig.~\ref{experiment_map}(b).%
Each mixture recording lasted more than 30~sec. and was windowed with 1~sec to make 30 samples for each location.
The evaluation was performed by leave-one-out cross-validation with the following procedure.
A mixture of samples at one location was chosen for evaluation and used the remaining samples for training.

The hyperparameters of NMF and GP were heuristically determined as follows.
The numbers of the basis vector for NMF were set to $L_k=5$ ($k=1, \ldots, K$) and $L_0=4$. 
The multiplication updates of NMF were iterated 100 times for both the training and localization.
The model parameters of GP regression were optimized by an Adam optimizer with a learning rate of 0.1 for 100 iterations.
The kernel parameter $\gamma_k$ was constrained to be greater than 3~m to represent the smooth spatial distribution of sound energy.
The localization was performed by a grid search on the likelihood function or posterior distribution.
The grid size was 0.1~m for $x$ and $y$ in the room, whereas a height of $z$ = 1~m was given; this is because a microphone attached to a human is usually at a consistent height.

The proposed method was evaluated under different signal-to-noise ratios (SNR).
A sound from the MIMII dataset\cite{Purohit2019} was added as out-of-domain background noise.
SNR was evaluated from -60 to +18~dB at every 3~dB. 

The advantage of the proposed method is a straightforward integration with other Bayesian localization methods.
To show the integration example, an integration with IMU-based localization was simulated.
IMU-based localization suffers from drift over time, while it works without any environmental setup.
The following probability distribution with location drift was chosen as an IMU-like prior:
\begin{align}
p(\x@) = \distnormal{(x, y) \mid (x_{\rm gt}, y_{\rm gt}) + (\epsilon_x, \epsilon_y)}{5^2\eye}
\end{align}
where $\epsilon_x, \epsilon_y = (5, 5)$ (m) %
are the drift components to simulate the drifted prior of IMU.
The expected localization error of $p(\x@)$ is approximately 7~m.

The proposed method was compared with na\"ive implementation of feature extraction and regression.
A feature extraction baseline with the Mel-frequency cepstrum coefficient (MFCC) with 20 components and a GP regression baseline were chosen to directly predict $x$ and $y$ coordinates of the microphone.
Localization results were evaluated with circular error (CE) and its empirical cumulative distribution function (eCDF).
A CE is a Euclidean distance between the true location and the estimated location in two dimensions.
These metrics are defined in ISO/IEC 18305:2016 and have been used for various indoor localization evaluations~\cite{ichikariOffSiteIndoorLocalization2019}.
The 50 percentiles (circular error probable: CEP), mean, and 95 percentiles (CE95) of CEs were used as representative values.

\begin{figure*}[!t]
    \centering
    \vspace{1mm}
    \includegraphics{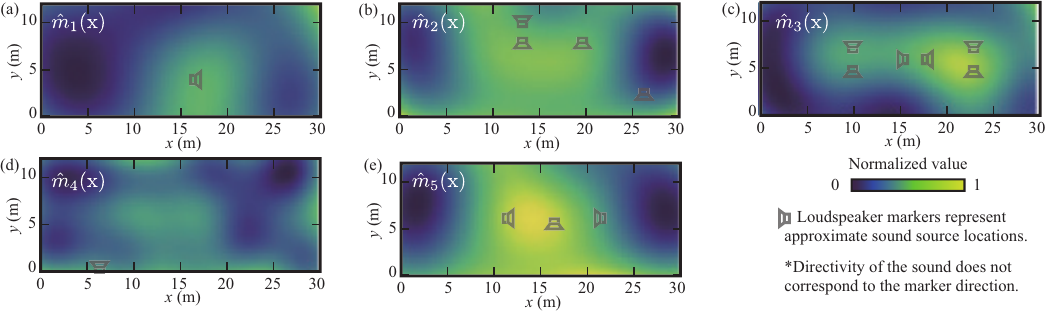}
    \vspace{-4mm}
    \caption{Examples of GP regression models with separated source RMS for each location. Figures correspond to the sound sources described in Fig.~\ref{experiment_map}: (a) a projector, (b) PCs, (c) air conditioners, (d) server room, and (e) ventilation fans, respectively.}
    \label{gp_map}
\end{figure*}

\subsection{Sound source separation and Gaussian regression}
Fig.~\ref{gp_map} shows examples of the predicted RMS (energy) of Gaussian process models.
The predicted RMS values have high values around the sound sources.
Slight directivity is also shown in Fig.~\ref{gp_map}(a).
The panel of the projector emitted environmental sound, which faced the positive $x$ direction.
The slight directivity is also observed in Fig.~\ref{gp_map}(c).
They may have vertical directivity since the RMS distribution is relatively narrow.
However, sounds from the PCs and ventilation fans in Figs.~\ref{gp_map}(b) and (e) diffract around the mechanical void at left and right.
One advantage of using supervised NMF is that it is easier to understand the features corresponding to the real world than nai\"ve spectral features.
From these results, we can confirm that these GP models provide reasonable RMS predictions.

\subsection{Localization performance}\label{multi-peak-discussion}

\begin{table}[t]
    \caption{Localization errors for each combination of the methods.}
    \label{location_errors}
    \centering
    \begin{tabular}{ll|ccc}
        \toprule
        \multicolumn{2}{c|}{Method} & \multicolumn{3}{c}{Localization error (m)} \\
        \midrule
        Feature   & Localization       & CEP & Mean & CE95 \\
        \midrule
        MFCC      & Regression         & 6.8 & 7.7 & 20.6 \\
        SNMF      & Regression         & 4.5 & 5.4 & 12.8 \\
        SNMF + WF & Regression         & 5.3 & 6.3 & 14.0 \\ %
        \midrule
        MFCC      & Likelihood         & 3.3 & 5.7 & 16.6 \\
        SNMF      & Likelihood         & 3.9 & 6.1 & 16.9 \\
        SNMF + WF & Likelihood         & 3.4 & 6.1 & 18.6 \\ %
        \midrule
        MFCC      & Likelihood + Prior & 3.0 & 5.1 & 15.2 \\
        SNMF      & Likelihood + Prior & 3.2 & 4.9 & 13.5 \\
        SNMF + WF & Likelihood + Prior & 2.7 & 3.3 &  9.2 \\
        \bottomrule
    \end{tabular}
    
\end{table}

\begin{figure}[!t]
    \centering
    \includegraphics{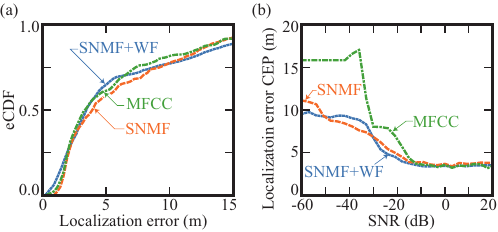}
    \caption{Localization performance. (a) Localization with proposed spatial likelihood in the experiment, (b) effect of SNR on localization error (CEP).}
    \label{ecdf}
\end{figure}

TABLE~\ref{location_errors} shows localization performance for all the combinations of feature extraction methods and regression methods.
The "SNMF" feature was the average temporal activation for each sound source $\h$ estimated by the supervised NMF.
The "SNMF-WF" feature (proposed) was the RMS of the extracted sound source (Eq.~\eqref{eq:feature}).
We first compare the performance differences of the direct regression and likelihood-based localization (top three rows vs. middle three rows).
Compared with the direct regression, the proposed spatial likelihood significantly reduced the localization errors in CEP.
The mean error and CE95, however, had no significant difference.
Since the proposed method first predicts the feature value at a location candidate and then compares it with the measurement, this process may create high peaks in the likelihood in completely different locations, which seem to have similar feature values (source energies).
Since environmental sounds may not be unique at several locations, the likelihood inherently becomes multi-modal distribution, and the microphone could not be uniquely localized from only audio data.
Fig.~\ref{ecdf}(a) show the localization performance using the proposed spatial-likelihood.
These results show that the proposed likelihood estimation is a reasonable extension of regression, with almost the same localization error.

\begin{figure*}[!t]
    \centering
    \vspace{1mm}
    \includegraphics{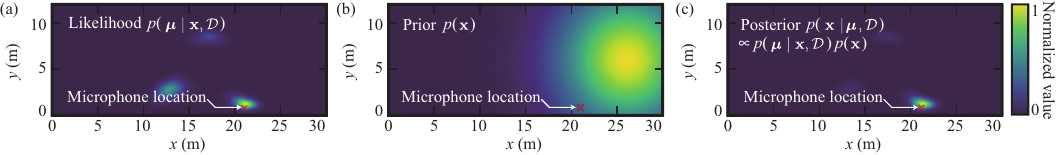}
    \vspace{-3mm}
    \caption{Example of integration with prior distribution. (a) spatial likelihood $p(\feat@ \mid \x@, \mathcal{D})$, (b) IMU-based prior distribution $p(\x@) = \distnormal{(x, y) \mid (26, 6)}{5^2I}$, and (c) posterior distribution $p(\x@ \mid \feat@, \mathcal{D}) \propto p(\feat@ \mid \x@, \mathcal{D})p(\x@)$ calculated from (a) and (b).}
    \label{posterior}
\end{figure*}

The effect of SNR on localization error is shown in Fig.~\ref{ecdf}(b).
There is a certain threshold according to SNR, beyond which the accuracy deteriorates rapidly.
The localization error of MFCC has a sharp change in accuracy starting at about -15~dB and saturating at around -24~dB. The relevant fingerprint sounds were mistaken for fingerprints from another location.
The localization error of SNMF changes from a similar SNR, but does not have as drastic an effect on positional accuracy as MFCC.
This might be due to the fact that MFCC directly uses mixture signals, which drastically change according to SNR, whereas SNMF and SNMF-WF first decompose the input mixture into individual source signals, which might not change significantly.
The localization error of the SNMF-based method saturates at around -33~dB, which is a 9~dB improvement from MFCC.
        
\subsection{Integration with other localization method}
Finally, the effectiveness of the proposed likelihood-based localization is demonstrated by comparing it with the na\"ive direct regression.
Simulating an IMU-based localization with large errors in its estimates, the posterior distribution was derived from our audio likelihood and the IMU-based prior to integrate these two modalities.
Fig.~\ref{posterior} shows an example of the integration results.
As mentioned in section~\ref{multi-peak-discussion}, the spatial likelihood can have multiple peaks at different locations.
Fig.~\ref{posterior}(a) shows such a case, where the likelihood has three sharp peaks.
Fig.~\ref{posterior}(b) shows a drifted prior distribution.
The combination of these distributions narrows the choice of location, as shown in Fig.~\ref{posterior}(c).
In this case, the localization error was reduced even with the biased prior, whose expected error is approximately 7~m, as shown in the bottom three rows of TABLE~\ref{location_errors}. %
Note that prior distribution does not always improve the accuracy, such as when the prior picks up a different peak of likelihood.
From the above example, we can confirm the straightforward Bayesian integration of the proposed method enables collaboration with existing methods.
Time-series tracking by particle filtering is also possible, which is in future work.

\section{Conclusions}
In this study, a method for infrastructure-free indoor microphone localization from indoor environmental sounds was proposed.
The proposed method first extracts landmark sounds from an observed a mixture signal using supervised NMF and Wiener filtering.
Then, the microphone is localized by maximizing the spatial likelihood from the GP regression.
The proposed method was evaluated in an approximately 12~m $\times$ 30~m room with five types of environmental sounds.
The results showed that NMF-based feature extraction improves the localization performance, specifically against noise from the na\"ive MFCC feature.
The likelihood-based localization was demonstrated to be easily integrated with a prior distribution.
This characteristic enables the further extension of the proposed method with sensor fusion.

Future work should involve time-series estimation and fusion with a magnetic or IMU-based method.
Since the proposed method uses the RMS of each sound, sound propagation simulation can generate the fingerprint.

\section*{Acknowledgments}
This work was supported in part by JSPS KAKENHI under Grant JP22K17922 and JST ACT-X No. JPMJAX200N.

\bibliographystyle{IEEEtran}
\bibliography{references.bib}

\end{document}